\pdfoutput=1
 \documentclass[aps,prd,showpacs,superscriptaddress,letterpaper,twocolumn,preprintnumbers]{revtex4-1}

\usepackage{dcolumn}

\usepackage{color}
\usepackage{eso-pic}

\newif\ifPDF
 \ifx\pdfoutput\undefined
 \pdffalse                  
\else
 \pdfoutput=1               
 \PDFtrue
\fi

\ifPDF
 \usepackage{url,boxedminipage}
 \usepackage{graphicx,thumbpdf}
 \definecolor{rltred}{rgb}{0.75,0,0}
 \definecolor{rltgreen}{rgb}{0,0.5,0}
 \definecolor{rltblue}{rgb}{0,0,0.75}
 \usepackage[colorlinks%
  ,bookmarks%
  ,urlcolor=rltblue%
  ,filecolor=rltgreen%
  ,linkcolor=rltred%
  ,pdftitle={pdftitLe}%
  ,pdfsubject={pdfsubject}%
  ,pdfkeywords={pdfkeywords}%
  ,pdfauthor={Joe User <joeuser@fnal.gov>}%
  ,pdfpagemode={UseOutlines}%
  ,bookmarksopen=true%
  ,bookmarksnumbered=true%
  ,pdfstartview={Fit}%
  ]{hyperref}
 \DeclareGraphicsExtensions{.pdf,.jpg,.jpeg}
\else
 \usepackage{graphicx}
 \DeclareGraphicsExtensions{.eps,.ps,.eps.gz,.ps.gz}
 \newcommand{\href}[2]{#2}                   
 \usepackage[dvips,pagebackref]{hyperref}    
\fi 

\usepackage{dcolumn}
\usepackage{bm}

\newcommand{\Zee}{\ensuremath{Z \rightarrow e^+e^-}}

\newcommand{\ppbar}{\ensuremath{p\bar{p}}}
\newcommand{\Znn}{\ensuremath{Z \rightarrow \nu \bar{\nu}}}
\newcommand{\Znng}{\ensuremath{Z\gamma \rightarrow \nu \bar{\nu}\gamma}}
\newcommand{\Zmm}{\ensuremath{Z \rightarrow \mu^+ \mu^-}}

\newcommand{\Zg}{\ensuremath{Z\gamma}}

\newcommand{\GeV}{\ensuremath{\mathrm{Ge\kern -0.1em V}}}
\newcommand{\GeVc}{\ensuremath{\GeV\!/c}}
\newcommand{\GeVcc}{\ensuremath{\GeV\!/c^2}}

\newcommand{\invfb}{\ensuremath{\mathrm{fb^{-1}}}}

\newcommand{\Pt}{\ensuremath {p_{{T}}}}
\def\sla#1{\rlap{\kern .15em /}#1}
\newcommand{\met}{\ensuremath{\sla{E}_T}}
\newcommand{\Et}{\ensuremath{E_{T}}}
\newcommand{\Etg}{\ensuremath{E_{T}^{\gamma}}}

\begin{document}

%

\preprint{CDF/PUB/ELECTROWEAK/CDFR/9904}
\preprint{FERMILAB-PUB-10-073-E}

\title{
Measurement of {\boldmath $Z\gamma$} Production in 
{\boldmath $p\overline{p}$} Collisions at
{\boldmath $\sqrt{s}$} = 1.96 TeV\\
}
\font\eightit=cmti8
\def\r#1{\ignorespaces $^{#1}$}

\affiliation{Institute of Physics, Academia Sinica, Taipei, Taiwan 11529, Republic of China} 
\affiliation{Argonne National Laboratory, Argonne, Illinois 60439, USA} 
\affiliation{University of Athens, 157 71 Athens, Greece} 
\affiliation{Institut de Fisica d'Altes Energies, Universitat Autonoma de Barcelona, E-08193, Bellaterra (Barcelona), Spain} 
\affiliation{Baylor University, Waco, Texas 76798, USA} 
\affiliation{Istituto Nazionale di Fisica Nucleare Bologna, $^{dd}$University of Bologna, I-40127 Bologna, Italy} 
\affiliation{Brandeis University, Waltham, Massachusetts 02254, USA} 
\affiliation{University of California, Davis, Davis, California 95616, USA} 
\affiliation{University of California, Los Angeles, Los Angeles, California 90024, USA} 
\affiliation{University of California, San Diego, La Jolla, California 92093, USA} 
\affiliation{University of California, Santa Barbara, Santa Barbara, California 93106, USA} 
\affiliation{Instituto de Fisica de Cantabria, CSIC-University of Cantabria, 39005 Santander, Spain} 
\affiliation{Carnegie Mellon University, Pittsburgh, Pennsylvania 15213, USA} 
\affiliation{Enrico Fermi Institute, University of Chicago, Chicago, Illinois 60637, USA}
\affiliation{Comenius University, 842 48 Bratislava, Slovakia; Institute of Experimental Physics, 040 01 Kosice, Slovakia} 
\affiliation{Joint Institute for Nuclear Research, RU-141980 Dubna, Russia} 
\affiliation{Duke University, Durham, North Carolina 27708, USA} 
\affiliation{Fermi National Accelerator Laboratory, Batavia, Illinois 60510, USA} 
\affiliation{University of Florida, Gainesville, Florida 32611, USA} 
\affiliation{Laboratori Nazionali di Frascati, Istituto Nazionale di Fisica Nucleare, I-00044 Frascati, Italy} 
\affiliation{University of Geneva, CH-1211 Geneva 4, Switzerland} 
\affiliation{Glasgow University, Glasgow G12 8QQ, United Kingdom} 
\affiliation{Harvard University, Cambridge, Massachusetts 02138, USA} 
\affiliation{Division of High Energy Physics, Department of Physics, University of Helsinki and Helsinki Institute of Physics, FIN-00014, Helsinki, Finland} 
\affiliation{University of Illinois, Urbana, Illinois 61801, USA} 
\affiliation{The Johns Hopkins University, Baltimore, Maryland 21218, USA} 
\affiliation{Institut f\"{u}r Experimentelle Kernphysik, Karlsruhe Institute of Technology, D-76131 Karlsruhe, Germany} 
\affiliation{Center for High Energy Physics: Kyungpook National University, Daegu 702-701, Korea; Seoul National University, Seoul 151-742, Korea; Sungkyunkwan University, Suwon 440-746, Korea; Korea Institute of Science and Technology Information, Daejeon 305-806, Korea; Chonnam National University, Gwangju 500-757, Korea; Chonbuk National University, Jeonju 561-756, Korea} 
\affiliation{Ernest Orlando Lawrence Berkeley National Laboratory, Berkeley, California 94720, USA} 
\affiliation{University of Liverpool, Liverpool L69 7ZE, United Kingdom} 
\affiliation{University College London, London WC1E 6BT, United Kingdom} 
\affiliation{Centro de Investigaciones Energeticas Medioambientales y Tecnologicas, E-28040 Madrid, Spain} 
\affiliation{Massachusetts Institute of Technology, Cambridge, Massachusetts 02139, USA} 
\affiliation{Institute of Particle Physics: McGill University, Montr\'{e}al, Qu\'{e}bec, Canada H3A~2T8; Simon Fraser University, Burnaby, British Columbia, Canada V5A~1S6; University of Toronto, Toronto, Ontario, Canada M5S~1A7; and TRIUMF, Vancouver, British Columbia, Canada V6T~2A3} 
\affiliation{University of Michigan, Ann Arbor, Michigan 48109, USA} 
\affiliation{Michigan State University, East Lansing, Michigan 48824, USA}
\affiliation{Institution for Theoretical and Experimental Physics, ITEP, Moscow 117259, Russia} 
\affiliation{University of New Mexico, Albuquerque, New Mexico 87131, USA} 
\affiliation{Northwestern University, Evanston, Illinois 60208, USA} 
\affiliation{The Ohio State University, Columbus, Ohio 43210, USA} 
\affiliation{Okayama University, Okayama 700-8530, Japan} 
\affiliation{Osaka City University, Osaka 588, Japan} 
\affiliation{University of Oxford, Oxford OX1 3RH, United Kingdom} 
\affiliation{Istituto Nazionale di Fisica Nucleare, Sezione di Padova-Trento, $^{ee}$University of Padova, I-35131 Padova, Italy} 
\affiliation{LPNHE, Universite Pierre et Marie Curie/IN2P3-CNRS, UMR7585, Paris, F-75252 France} 
\affiliation{University of Pennsylvania, Philadelphia, Pennsylvania 19104, USA}
\affiliation{Istituto Nazionale di Fisica Nucleare Pisa, $^{ff}$University of Pisa, $^{gg}$University of Siena and $^{hh}$Scuola Normale Superiore, I-56127 Pisa, Italy} 
\affiliation{University of Pittsburgh, Pittsburgh, Pennsylvania 15260, USA} 
\affiliation{Purdue University, West Lafayette, Indiana 47907, USA} 
\affiliation{University of Rochester, Rochester, New York 14627, USA} 
\affiliation{The Rockefeller University, New York, New York 10021, USA} 
\affiliation{Istituto Nazionale di Fisica Nucleare, Sezione di Roma 1, $^{ii}$Sapienza Universit\`{a} di Roma, I-00185 Roma, Italy} 

\affiliation{Rutgers University, Piscataway, New Jersey 08855, USA} 
\affiliation{Texas A\&M University, College Station, Texas 77843, USA} 
\affiliation{Istituto Nazionale di Fisica Nucleare Trieste/Udine, I-34100 Trieste, $^{jj}$University of Trieste/Udine, I-33100 Udine, Italy} 
\affiliation{University of Tsukuba, Tsukuba, Ibaraki 305, Japan} 
\affiliation{Tufts University, Medford, Massachusetts 02155, USA} 
\affiliation{Waseda University, Tokyo 169, Japan} 
\affiliation{Wayne State University, Detroit, Michigan 48201, USA} 
\affiliation{University of Wisconsin, Madison, Wisconsin 53706, USA} 
\affiliation{Yale University, New Haven, Connecticut 06520, USA} 
\author{T.~Aaltonen}
\affiliation{Division of High Energy Physics, Department of Physics, University of Helsinki and Helsinki Institute of Physics, FIN-00014, Helsinki, Finland}
\author{J.~Adelman}
\affiliation{Enrico Fermi Institute, University of Chicago, Chicago, Illinois 60637, USA}
\author{B.~\'{A}lvarez~Gonz\'{a}lez$^w$}
\affiliation{Instituto de Fisica de Cantabria, CSIC-University of Cantabria, 39005 Santander, Spain}
\author{S.~Amerio$^{ee}$}
\affiliation{Istituto Nazionale di Fisica Nucleare, Sezione di Padova-Trento, $^{ee}$University of Padova, I-35131 Padova, Italy} 

\author{D.~Amidei}
\affiliation{University of Michigan, Ann Arbor, Michigan 48109, USA}
\author{A.~Anastassov}
\affiliation{Northwestern University, Evanston, Illinois 60208, USA}
\author{A.~Annovi}
\affiliation{Laboratori Nazionali di Frascati, Istituto Nazionale di Fisica Nucleare, I-00044 Frascati, Italy}
\author{J.~Antos}
\affiliation{Comenius University, 842 48 Bratislava, Slovakia; Institute of Experimental Physics, 040 01 Kosice, Slovakia}
\author{G.~Apollinari}
\affiliation{Fermi National Accelerator Laboratory, Batavia, Illinois 60510, USA}
\author{J.~Appel}
\affiliation{Fermi National Accelerator Laboratory, Batavia, Illinois 60510, USA}
% ...

\collaboration{CDF Collaboration\footnote{With visitors from $^a$University of Massachusetts Amherst, Amherst, Massachusetts 01003,
$^b$Universiteit Antwerpen, B-2610 Antwerp, Belgium, 
$^c$University of Bristol, Bristol BS8 1TL, United Kingdom,
$^d$Chinese Academy of Sciences, Beijing 100864, China, 
$^e$Istituto Nazionale di Fisica Nucleare, Sezione di Cagliari, 09042 Monserrato (Cagliari), Italy,
$^f$University of California Irvine, Irvine, CA  92697, 
$^g$University of California Santa Cruz, Santa Cruz, CA  95064, 
$^h$Cornell University, Ithaca, NY  14853, 
$^i$University of Cyprus, Nicosia CY-1678, Cyprus, 
$^j$University College Dublin, Dublin 4, Ireland,
$^k$University of Edinburgh, Edinburgh EH9 3JZ, United Kingdom, 
$^l$University of Fukui, Fukui City, Fukui Prefecture, Japan 910-0017,
$^m$Kinki University, Higashi-Osaka City, Japan 577-8502,
$^n$Universidad Iberoamericana, Mexico D.F., Mexico,
$^o$University of Iowa, Iowa City, IA  52242,
$^p$Kansas State University, Manhattan, KS 66506,
$^q$Queen Mary, University of London, London, E1 4NS, England,
$^r$University of Manchester, Manchester M13 9PL, England,
$^s$Muons, Inc., Batavia, IL 60510, 
$^t$Nagasaki Institute of Applied Science, Nagasaki, Japan, 
$^u$University of Notre Dame, Notre Dame, IN 46556,
$^v$Obninsk State University, Obninsk, Russia,
$^w$University de Oviedo, E-33007 Oviedo, Spain, 
$^x$Texas Tech University, Lubbock, TX  79609, 
$^y$IFIC(CSIC-Universitat de Valencia), 56071 Valencia, Spain,
$^z$Universidad Tecnica Federico Santa Maria, 110v Valparaiso, Chile,
$^{aa}$University of Virginia, Charlottesville, VA  22906,
$^{bb}$Bergische Universit\"at Wuppertal, 42097 Wuppertal, Germany,
$^{cc}$Yarmouk University, Irbid 211-63, Jordan,
$^{kk}$On leave from J.~Stefan Institute, Ljubljana, Slovenia, 
}}
\noaffiliation

\date{\today}

\begin{abstract}
The production rate and kinematics of photons produced in association with $Z$ bosons are 
studied using 2 \invfb\ of \ppbar\ collision data collected at the Collider 
Detector at Fermilab.  
The cross section for $\ppbar\to \ell^+\ell^-\gamma + X$ (where the leptons $\ell$ are either muons or electrons with dilepton mass  $M_{\ell\ell} > 40\, \GeVcc$, and where the photon has transverse energy $\Etg > 7$ GeV and is well separated from the leptons)
is 4.6 $\pm$ 0.2\,(stat) $\pm$ 0.3\,(syst) $\pm$ 0.3\,(lum) pb, which is consistent with standard model expectations.  
We use the photon \Et\ distribution from \Zg\ events where the $Z$ has decayed to $\mu^+\mu^-$, $e^+e^-$, or $\nu\bar{\nu}$ to set limits on anomalous (non-standard-model) trilinear couplings between photons and $Z$ bosons.
\end{abstract}

\pacs{PACS numbers 14.70.Bh, 14.70.Hp, 13.85.Qk}

\maketitle

The study  of  \Zg\ production~\cite{Znotation} is an important test of the 
standard model (SM)
description of gauge boson interactions and provides sensitivity to physics beyond the SM~\cite{TNGBC2001}.
The \Zg\  cross section is directly sensitive to the trilinear gauge couplings at the $ZZ\gamma$ and $Z\gamma\gamma$ vertices, which are predicted to vanish in the SM at tree level~\cite{tgc,prd41_1990_1476,prd47_1993_4889}.
Physics beyond the SM (e.g.\ compositeness or supersymmetry)
could alter the cross section and the production kinematics. 
\Zg\ production is also an important background 
to searches for new physics (e.g.~in Gauge-Mediated Supersymmetry Breaking models~\cite{gmsb}) and Higgs searches.
In this paper the production properties of  \Zg\ events are compared 
to SM predictions, and limits are set on anomalous trilinear couplings.

We present measurements of  $Z \gamma$  production from 
\ppbar\ collisions at $\sqrt{s}$ = 1.96 TeV at the Tevatron Collider using data 
obtained with the Collider Detector at Fermilab (CDF II). 
We use two methods to identify \Zg\ events.  For events where the $Z$ decays to charged leptons, we first identify the $Z$ decay products and then look for the associated photon~\cite{photonID}.  We use these events to measure the \Zg\ cross section since we are able to identify them with low backgrounds even at low photon transverse energy (\Etg)~\cite{eta_defn}.  Anomalous (non-SM) \Zg\ couplings (described below) would produce an excess of events with high \Etg, so to set limits on these couplings we also include events where the $Z$ boson decays into neutrinos.  We identify these by looking for events with high \Etg\  and large missing transverse energy (\met)~\cite{MET_defn}. 
An integrated luminosity of $2.0$ fb$^{-1}$ is used for
$Z$ bosons selected in the \Zmm\ and \Znn\ decay modes and 1.1 \invfb\ is used for the \Zee\ decay mode.


The CDF detector is described in detail elsewhere~\cite{CDF,CDF_run2}. 
The transverse momenta (\Pt) of charged particles
are measured by an eight-layer silicon strip detector~\cite{SVX} 
and a 96-layer drift chamber (COT)~\cite{COT} inside a 1.4 Tesla magnetic field.
The COT provides tracking coverage with high efficiency for $|\eta|<1$. 
Electromagnetic and hadronic calorimeters
surround the tracking system. They are segmented in a projective tower geometry and measure
the energies of charged and neutral particles in the central ($|\eta|<1.1$) and forward ($1.1<|\eta|<3.6$) regions. 
Each calorimeter has an electromagnetic shower profile detector 
positioned at the shower maximum~\cite{showerMax}.  
The calorimeters are surrounded by muon drift chambers 
covering $|\eta|<1$.
Gas Cherenkov counters~\cite{CLC} measure the average number of $p\bar{p}$ 
inelastic collisions per beam crossing and thereby determine the beam luminosity.
The readout decision is made with a 
fast three-level trigger system that has high efficiency for
selecting the $Z$ bosons to be used in the offline 
analysis.

For the \Zg\ cross section measurement, we select $Z\gamma$ candidate events by identifying \Zmm\ and \Zee\ and then looking for photon candidates with $\Etg > 7$ GeV. Event selection starts with inclusive muon (electron) triggers requiring muon $\Pt > 18$ \GeVc\ (electron $\Et > 18 $ GeV). 
Further selection requires the trigger muon (electron) to be isolated and to have $\Pt > 20$ \GeVc\ ($\Et > 20$ GeV) and to pass standard muon (electron) identification criteria.
Once we have identified a trigger lepton, we require a second oppositely charged lepton of the same flavor.   For the second muon, we accept an isolated track with $\Pt > 20$ \GeVc, $|\eta| < 1$, with a minimum-ionizing energy deposit in the calorimeter, regardless of whether or not it is associated with a reconstructed track in the muon chambers.  For the second electron, we accept isolated electromagnetic clusters with $\Et > 20$ GeV and $|\eta| < 2.8$.  One of the electrons is allowed to pass a looser isolation requirement, with up to three times the amount of additional energy near the electron compared to the standard isolation requirement.  The shape of the shower in the shower max detector is required to match the expected shape for an electron, and in the central region we require a match between this shower and a track.  Since the tracking efficiency falls off in the forward region ($|\eta| > 1.2$), we do not require forward electrons to be matched to a track and therefore we do not require opposite charge.  The dilepton invariant mass for both $\mu^+\mu^-$ and $e^+e^-$ is required to be larger than 40 \GeVcc.

We measure trigger and reconstruction efficiencies by studying leptonic decays of $W$ and $Z$ bosons. 
We measure muon trigger efficiencies using \Zmm\  decays where the two muons are fiducial to different muon detector systems.  The muon triggers varied during the running period;  most of the data were taken with triggers that only required a track match in the $r$--$\phi$ plane while a 3-dimensional track match was required later during high-luminosity running.  The average trigger efficiency for muons is $(91.6 \pm 1.0)$\%.  We also use the two muons from \Zmm\ decays to measure the average 
muon reconstruction and identification efficiency to be $(81.3 \pm 1.1)$\%.

 To find the trigger efficiency for a central electron that satisfies the selection criteria, we use $W$ boson candidates triggered by an electromagnetic cluster and \met\ but with no tracking requirement.  The trigger efficiency for electrons passing the final selection is nearly 100\% for electrons with a high-\Pt\ track ($> 80$ \GeVc), averages $(95.9\pm0.6)$\% for electrons with track \Pt\ between 20 and 80 \GeVc, and falls sharply for electrons with track $\Pt < 14$ \GeVc.
We also use the two electrons of $Z$ boson candidates to measure the electron reconstruction and identification efficiency.
This falls slightly during the run as the luminosity increases; for central electrons it is in the range of 95\% to 96\% and for forward electrons it is in the range of 90\% to 92\%.

A small fraction of the $Z$ boson candidates are background, primarily from QCD jets that are reconstructed as electrons.  In the central region we use same-sign events to estimate the number of background events, which is found to be negligible in both the muon and the electron datasets.
In events with one central electron and one forward electron, the charge of the forward electron is not measured 
reliably, so we normalize to the high-mass sideband and find $(7.7 \pm 1.0)$\% background.

Once we have selected events with $Z$ boson candidates, we look for well-identified isolated~\cite{photonIsolation} photons in the central region $(|\eta| < 1.1)$ with $\Etg > 7$ GeV that are well separated from the $Z$ decay leptons $(\Delta R_{\ell\gamma} > 0.7)$.  To determine the efficiency for identifying real photons, we select a sample of $Z$ events where one of the final-state leptons has radiated a photon.  We use tight lepton selection criteria to reduce QCD backgrounds and very loose criteria for selecting the photon to insure nearly full acceptance for photons that have not converted inside the tracking chamber.  We reduce the background fraction to a negligible level by requiring the three-body invariant mass of the two leptons and the photon to be $91 \pm 5$ \GeVcc.  Using this method we determine that the efficiency for identifying real photons is $(86 \pm 2)$\%, with the largest inefficiency coming from the isolation requirement, which is needed to reduce QCD backgrounds.

The largest background to the photons comes from QCD jets.  A jet can fragment into a high-\Pt\ $\pi^0$ which decays into two photons.  If the two photons are sufficiently collinear, their electromagnetic showers in the calorimeter will be indistinguishable from a single-photon shower.   To estimate the number of QCD background events in the $Z\gamma$ sample we use the jets in the $Z$ sample and a probability function for jets to fragment in such a way that they pass the photon identification criteria.  This probability function is determined from events collected by a dijet trigger by measuring the rate at which we identify photons (as a function of jet \Et) and subtracting the expected true photon production rate.  We also correct for the difference in the ratio of jets coming from quarks versus gluons, since quark jets are much more likely to be reconstructed as photons. 
The probability is about 0.29\% at 10~GeV and drops rapidly to about 0.07\% (with a large systematic uncertainty) for jets above 25~GeV.  The uncertainty on the photon background is the dominant uncertainty on the $Z\gamma$ cross section measurement.

We find a total of 778 \Zg\ events: 390 $e^+e^-\gamma$ events in $1.07\, \pm\, 0.06$  \invfb\ and 388 $\mu^+\mu^-\gamma$ events in $2.01\pm 0.12$ \invfb.  We estimate the QCD photon background to be $94 \pm 26$ events (primarily at low photon \Et), and we estimate the number of events where a jet is misidentified as an electron to be $14\, \pm\, 7$ events.  This gives measured cross sections of $\sigma(\ppbar\to e^+e^-\gamma+X) = 4.9 \pm 0.3\, $(statistical) $ \pm 0.3\,$(systematic) $ \pm 0.3\,$(luminosity) pb and $\sigma(\ppbar\to \mu^+\mu^-\gamma+X) = 4.4 \pm 0.3 \pm 0.2 \pm 0.3$ pb.  The average cross section is  $\sigma(\ppbar\to \ell^+\ell^-\gamma+X) = 4.6\, \pm\, 0.2 \pm\, 0.3\pm\, 0.3$ pb, which agrees with a next-to-leading-order SM calculation~\cite{NLOcalc} of $4.5\, \pm\, 0.4$ pb.  These cross sections are for events with $\Etg > 7$ GeV, $\Delta R_{\ell\gamma} > 0.7$, and a dilepton mass of at least 40 \GeVcc.

\begin{figure}
\begin{center}
  \includegraphics[width=0.98 \columnwidth]{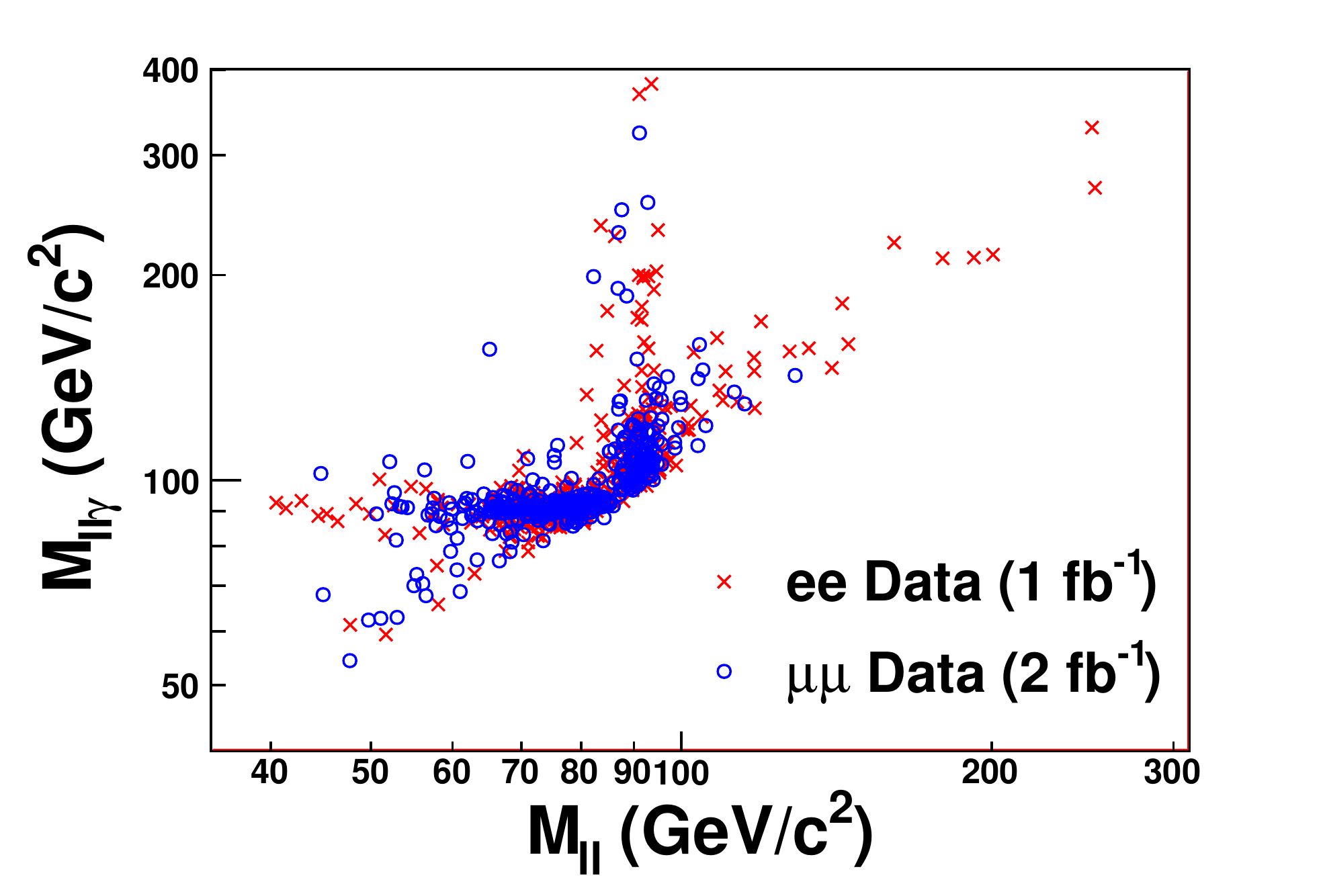}
  \caption{Three-body $M_{ll\gamma}$ vs. two-body {$M_{ll}$  in the \Zmm\ and \Zee\ datasets for photons with $\Et > 7$ GeV.}
  \label{fig:MllVsMllg1GeV_Zll}}
\end{center}
\end{figure}

	There is no direct coupling between the $Z$ boson and the photon in the SM, so $Z\gamma$ production at tree level comes from either initial-state radiation (ISR), where the photon is radiated by one of the incoming quarks, or final-state radiation (FSR), where the photon is radiated off one of the charged leptons from the decay of the $Z$ boson.  In ISR events the dilepton mass will approximate the $Z$ mass, whereas in FSR events the three-body lepton-lepton-photon mass will approximate the $Z$ mass.  In Fig.~\ref{fig:MllVsMllg1GeV_Zll} we plot the three-body mass on the vertical axis and the dilepton mass on the horizontal axis.  FSR-dominated events form the horizontal line near $M_{\ell\ell\gamma} = 91$ \GeVcc, and ISR-dominated events form the vertical line near $M_{\ell\ell} = 91$ \GeVcc.  The diagonal contains radiative Drell-Yan events.

\begin{figure}
   \begin{center}
       \includegraphics[width=0.98 \columnwidth]{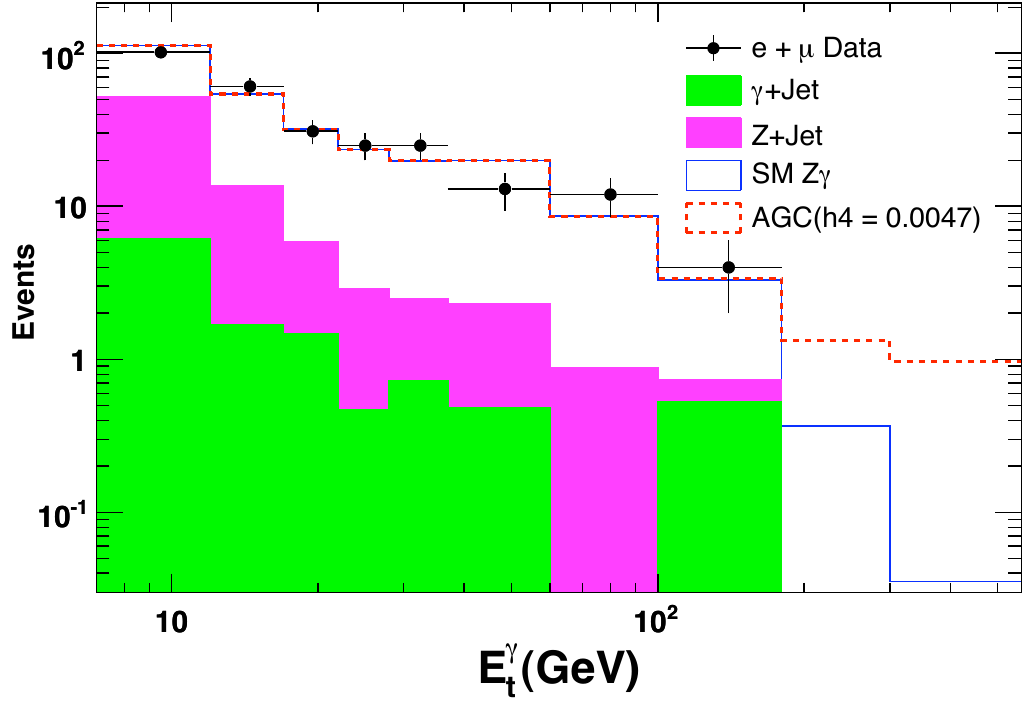}
        \caption{Photon \Et\ distribution for ISR-dominated  events ($M_{ll\gamma} > 100$ \GeVcc) in the $ l^+l^-\gamma$ dataset.  The solid line is the SM prediction, while the dotted line is the prediction including an anomalous coupling $h_4 = 0.0047$.  The predictions were made using the ZGAMMA~\cite{prd47_1993_4889} generator.}  
       \label{fig:ISREt}
   \end{center}    
   \end{figure}

Potential anomalous $Z\gamma$ couplings have been studied by Baur and Berger~\cite {prd47_1993_4889} and we adopt their notation for classifying the nature of the coupling: $h^V_i$ where $V$ is either a $Z$ or a $\gamma$ and the index $i$ runs from 1 to 4.  In order to preserve tree-level unitarity at large center-of-mass energy $\sqrt{\hat{s}}$, a form factor is used of the form $h^V_i = {{h^V_{i0}}\over{(1+\hat{s} /  \Lambda^2)^n}}$. In this paper we use $\Lambda = 1.2$ TeV and $n=3(4)$ for $i=3(4)$.
If an anomalous coupling between the $Z$ boson and the photon exists, it will produce more events with high-\Etg\ photons than expected from the SM.  An example of this is included in  Fig.~\ref{fig:ISREt}, which compares the 
distribution of \Etg\ for ISR events (events where $M_{ll\gamma} > 100$ \GeVcc) to the SM expectation, showing good agreement.

	To set limits on anomalous $Z\gamma$ couplings, we use the same dataset used for the cross section measurement. Since the statistics are limited at high \Etg, we also include \Znng\ events identified from a photon plus \met\ dataset where $\Etg > 90$ GeV.  The photon plus \met\ dataset used for this analysis is identical to that used in~\cite{gammaMet2008}.  

\begin{figure}
   \begin{center}
       \includegraphics[width=0.98\columnwidth]{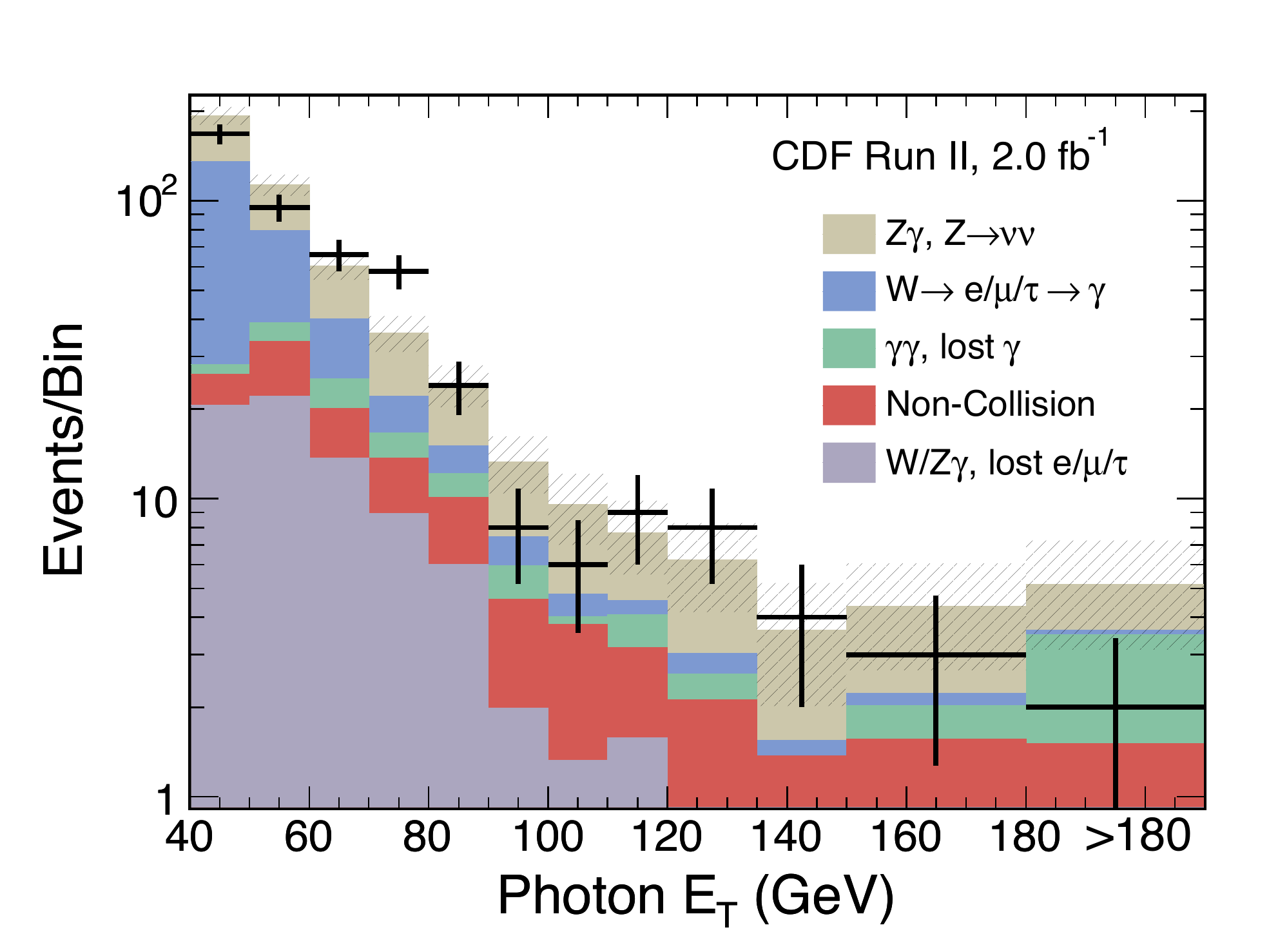}
        \caption{Photon \Et\ distribution for  \Znng\ candidate events.  The data are compared to SM contributions. Only events with $\Etg > 90$ GeV are used for this analysis. The cross hatching represents the uncertainty on the predicted number of events  (from~\cite{gammaMet2008}).}  
       \label{fig:phoEt}
   \end{center}    
   \end{figure}

	The expected signal and backgrounds for \Znng\ are described in detail in~\cite{gammaMet2008}.  
The total number of $\gamma\, +\, \met$ events with $\Etg > 90$ GeV is predicted to be $46.3\, \pm\, 3.0$, of which $25.6 \pm 2.0$ are expected to be \Znng\ events; we observe 40 events.  The \Etg\ distribution is displayed in Fig.~\ref{fig:phoEt} along with the predicted distributions for the signal and backgrounds.

	Using the measured \Etg\ distributions (separated by $Z$ decay channel) we set upper limits on the strength of anomalous couplings using a binned likelihood method. We determine the expected \Etg\ distributions from Monte Carlo data samples with non-zero anomalous couplings as well as SM couplings. 
There are too many samples to fully simulate each one, so instead we determine the net efficiency for a generated event to appear in the analysis sample as a function of \Etg\ by fully simulating both SM and representative anomalous coupling samples.  While the difference in efficiency between these samples is consistent with statistical fluctuations, we take the difference between the measured efficiencies as a systematic uncertainty.  We then apply this efficiency function to generator-level MC samples to get the expected \Etg\ distributions.
	
	The resulting upper limits on the strength of anomalous couplings are shown in Table~\ref{tab:AGClimits}.  The expected limits are determined by generating a large number of pseudo-experiments, Poisson smearing the number of expected signal and background events, and varying parameters subject to systematic uncertainties (which are dominated by uncertainties on the background estimates)~\cite{JianrongThesis}.  The observed limits are better than the average expected limits but fall within the expected range.  We see no evidence for anomalous couplings.  
The limits based upon events with \Zee\ and \Zmm\ decays are nearly identical to previous limits from the D0 collaboration~\cite{D0AGC1invfb}, which studied the same channels using 1 \invfb\ of data.  Our limits including the \Znn\ decays from 2~\invfb\ are not directly comparable to the limits published by the D0 collaboration~\cite{D0ZnunuAGC} using 3.6~\invfb\ since the D0 limits were calculated using $\Lambda = 1.5$ TeV.    

	In conclusion, we have measured the \Zg\ cross section at the Tevatron Collider and find that it is consistent with SM expectations.  We have also found that the \Etg\ distribution of photons produced in association with $Z$ bosons is consistent with SM couplings where there is no direct \Zg\ coupling, and we have set limits on anomalous gauge couplings.

    \begin{table}
    \begin{center}
    \begin{tabular}{rllc}
 \hline
\multicolumn{2}{l}{Observed Limit\hspace*{1 em}} & Expected Limit $\,$& Limit ($\mu+e$)\\
 \hline
           $|h_3^Z|$         &  $ < 0.050 $      & $0.062 \pm 0.014$            & $0.083$                     \\
           $|h_4^Z|$         &  $ < 0.0034  $   & $0.0043 \pm 0.0009$        & 0.0047                    \\
 \hline                                                                                            
           $|h_3^{\gamma}|$  &  $ < 0.051    $    & $0.064 \pm 0.014$      & 0.084              \\
           $|h_4^{\gamma}|$  &  $ < 0.0034  $    & $0.0043 \pm 0.0009$   & 0.0047            \\
 \hline
    \end{tabular}
       \caption{Upper limits (95\% C.L.) on anomalous $Z\gamma$ couplings using notation from reference~\cite{prd47_1993_4889} and $\Lambda=1.2$ TeV.  The last column gives the limit obtained using only the $\mu^+\mu^-\gamma$ and $e^+e^-\gamma$ events. 
     \label{tab:AGClimits}}  
    \end{center}
    \end{table}

We thank the Fermilab staff and the technical staffs of the participating 
institutions for their vital contributions. This work was supported by the 
U.S. Department of Energy and National Science Foundation; the Italian 
Istituto Nazionale di Fisica Nucleare; the Ministry of Education, Culture, 
Sports, Science and Technology of Japan; the Natural Sciences and 
Engineering Research Council of Canada; the National Science Council of the 
Republic of China; the Swiss National Science Foundation; the A.P. Sloan 
Foundation; the Bundesministerium f\"ur Bildung und Forschung, Germany; the 
World Class University Program, the National Research Foundation of Korea; 
the Science and Technology Facilities Council and the Royal 
Society, UK; the Institut National de Physique Nucleaire et Physique des 
Particules/CNRS; the Russian Foundation for Basic Research; the Ministerio 
de Ciencia e Innovaci\'{o}n, and Programa Consolider-Ingenio 2010, Spain; the 
Slovak R\&D Agency; and the Academy of Finland.


\begin{thebibliography}{99}
\bibitem{Znotation}
Throughout this paper the notation ``$Z$'' is used to specify $Z/\gamma^*$ production via the Drell-Yan process.
\bibitem{eta_defn} We use a coordinate system where $\phi$ is the
  azimuthal angle around the beam axis and $\theta$ is the polar angle 
  measured with respect to the proton direction.
  The pseudo-rapidity $\eta$ is defined by $\eta \equiv -\ln\tan(\theta/2)$, 
  and $\Delta R \equiv \sqrt{ \Delta\eta^2 + \Delta\phi^2 }$.
\bibitem{MET_defn}
  Transverse momentum and energy are defined as $\Pt = p \sin \theta$ and $\Et = E \sin \theta$, respectively. Missing \Et\  (\met ) is defined by $\vec{\met} = -\Sigma_i \Et^i\hat{n}_i$, where $i$ is the calorimeter tower number for 
$|\eta | < 3.6$, and $\hat{n}_i$ is a  unit vector perpendicular to the beam axis and pointing at the $i^{th}$ tower. ($\met = |\vec{\met}|$)
\bibitem{photonIsolation} The additional energy in a cone of $\Delta R < 0.4$ must be less than $ 0.1 \times \Etg$ if $\Etg < 20$ GeV and less than $2+0.02\times(\Etg - 20)$ if $\Etg > 20$ GeV for a photon to pass the isolation selection.
\end{thebibliography}
\end{document}